\documentclass{emulateapj}

\shorttitle{Super-critical Accretion Flows around Black Holes}
\shortauthors{Ohsuga et al.}

\newcommand{\lsim}{\raisebox{0.3mm}{\em $\, <$} \hspace{-2.8mm}
\raisebox{-1.8mm}{\em $\sim \,$}}
\newcommand{\gsim}{\raisebox{0.3mm}{\em $\, >$} \hspace{-2.8mm}
\raisebox{-1.8mm}{\em $\sim \,$}}
\newcommand{\bm}[1]{\mbox{\boldmath $#1$}}

\begin{document}

\title{Super-critical Accretion Flows around Black Holes:
Two-dimensional, Radiation-pressure-dominated Disks with Photon-trapping}


\author{K. Ohsuga\altaffilmark{1,2}, 
M. Mori\altaffilmark{3,4}, 
T. Nakamoto\altaffilmark{5}, 
and S. Mineshige\altaffilmark{2}}


\altaffiltext{1}{Department of Physics, Rikkyo University, 
Toshimaku, Tokyo 171-8501, Japan}
\altaffiltext{2}{Yukawa Institute for Theoretical Physics,
Kyoto University, Kyoto 606-8502, Japan}
\altaffiltext{3}{Institute of Natural Sciences, Senshu University, 
Kawasaki, Kanagawa 214-8580, Japan}
\altaffiltext{4}{Department of Physics and Astronomy, 
University of California at Los Angeles,
Los Angeles, CA 90095-1562, USA}
\altaffiltext{5}{Center for Computational Sciences, University of Tsukuba, 
Tsukuba, Ibaraki 305-8577, Japan}


\begin{abstract}
The quasi-steady structure of super-critical accretion flows 
around a black hole is studied based on the
two-dimensional radiation-hydrodynamical (2D-RHD) simulations.
The super-critical flow is composed of two parts: 
the disk region and the outflow regions above and below the disk.
Within the disk region the circular motion as well as the patchy density
structure are observed, which is caused by Kelvin-Helmholtz instability
and probably by convection.
The mass-accretion rate decreases inward, roughly in proportion to the radius, 
and the remaining part of the disk material leaves the disk to form outflow 
because of strong radiation pressure force.
We confirm that photon trapping plays an important role within the disk.
Thus, matter can fall onto the black hole at a rate
exceeding the Eddington rate.
The emission is highly anisotropic and moderately collimated
so that the apparent luminosity can exceed the Eddington luminosity
by a factor of a few in the face-on view.
The mass-accretion rate onto the black hole 
increases with increase of the absorption opacity (metalicity) 
of the accreting matter.  This implies that 
the black hole tends to grow up faster in the metal rich regions 
as in starburst galaxies or star-forming regions.
\end{abstract}

\keywords{accretion: accretion disks --- black hole physics ---
hydrodynamics --- radiative transfer}

\section{INTRODUCTION}
It is widely believed that accretion flow onto black holes drives
major activities of astrophysical black holes,
such as active galactic nuclei (AGNs),
Galactic black hole candidates (BHCs), 
and possibly gamma-ray bursts (GRBs).
It is also a common belief that 
the basic accretion processes and radiation properties
can well be described by the standard-disk model 
by Shakura \& Sunyaev (1973).  On the other hand,
the observational facts which cannot fit the standard-disk 
picture are also being accumulated recently.
Good examples are intense high-energy emission from black holes, 
which indicates the presence of very hot plasmas around black holes
with temperature of $T \sim 10^9$K.  
This leads to the ideas of accretion disk corona
and/or radiatively inefficient flow (RIAF).

The standard-disk picture breaks down not only in the 
low-luminosity regimes but also in the high-luminosity regimes, 
in which mass-accretion rates, $\dot{M}$, becomes comparable to
or exceeds the critical mass-accretion rate,
$\dot{M}_{\rm crit} \equiv L_{\rm E}/c^2$,
with $L_{\rm E}$ being the Eddington luminosity.
Apparently, they look similar to those of the standard-type disks,
since the disk remains to be optically thick and thus emits blackbody-like 
emission, just as the standard-type disks do.
Flow dynamics is, however, distinct.

What makes the super-critical accretion flows distinct from
the standard-disk-type flow is the presence of photon trapping
(Begelman 1978).
Photon trapping occurs, when photon diffusion time, time
for photons to travel from the equatorial plane to the surface,
exceeds the accretion timescale.  
Under such circumstances photons generated via the viscous process
are advected inward with gas flow without being able to
go out from the surface immediately.
We can define the trapping radius, inside which photon trapping
is significant;
\begin{equation}
  r_{\rm trap} \sim \dot{m} \left( \frac{H}{r}\right) r_{\rm g},
  \label{rtrap}
\end{equation}
where 
$\dot{m}$ is the mass-accretion rate normalized by 
the critical mass-accretion rate, 
$H$ is the half thickness of the disk,
$r$ is the radius,
and $r_g (=2GM/c^2)$ is the Schwarzschild radius
with $M$ being the black hole mass
(Ohsuga et al. 2002, hereafter referred to as Paper I,
see also Begelman 1978 for the spherical case).

It is claimed that photon trapping effects were incorporated in the 
slim-disk model by Abramowicz et al. (1988;
for a review, see Kato, Fukue, \& Mineshige 1998, section 10.1).
We have shown previously, however, that
the slim-disk model does not fully consider
the photon trapping effects, since it is a
radially one-dimensional model, whereas the photon trapping 
is basically a multi-dimensional effect (see Paper I).
We thus need at least two-dimensional treatment.
Indeed, we have demonstrated in Paper I that 
the photon trapping is grossly underestimated in the slim-disk model.

Let us recall what lacks in the slim-disk model more explicitly.
The equation of energy balance, 
$Q_{\rm vis}=Q_{\rm rad}+Q_{\rm adv}$, is solved in the slim-disk model,
where $Q_{\rm vis}$, $Q_{\rm rad}$, and $Q_{\rm adv}$ 
are the viscous heating, radiative cooling, and 
advective cooling rates per unit surface, respectively.  
The problem resides in that
the radiative cooling is evaluated under the usual diffusion approximation
(in the vertical direction).
This approximation may be justified, 
if radial inflow of gas is totally negligible so that photons can
mainly diffuse in the vertical direction.
However, this is not the case, when
the diffusion timescale is longer than the accretion timescale.
This leads to over-estimation of $Q_{\rm rad}$ and, hence,
under-estimations of $Q_{\rm adv}$, compared with the correct value.
More importantly, photon trapping modifies spectral energy distribution (SED),
as well.  We have found that large photon trapping yields spectral softening,
because hard photons which are created deep inside the disk are more
effectively trapped than soft photons (Ohsuga, Mineshige, \& Watarai 2003).
Since our previous formulations are only partly multi-dimensional,
we, as a next step, need to perform fully two-dimensional analysis 
of the super-critical accretion flows.  This is a major motivation
of the present study.

When considering photon trapping effects, we should also
pay attention to the fact that the super-critical
accretion flow becomes geometrically thick.
The multi-dimensional gas motion, such as
convective or large-scale circulation, might occur.
Further, strong outflow might also be generated at the disk surface via
radiation pressure force.  
Such complex flow motion will influence radiation energy distribution 
through the advective energy transport, which, in turn, 
affects the flow motion via radiation pressure force. 
We need to carefully solve such strong coupling between radiation and matter.

Two-dimensional radiation-hydrodynamical (2D-RHD) simulations 
of accretion disks were initiated by Eggum, Coroniti, \& Katz (1987),
who assumed the equilibrium between gas and radiation.
The improved simulations, 
in which the energy of gas and radiation are separately treated,
were performed by Kley (1989), Okuda, Fujita, \& Sakashita (1997), 
Fujita \& Okuda (1998), Kley \& Lin (1999), and Okuda \& Fujita (2000).

The simulations of super-critical flows around black holes 
were pioneered by Eggum, Coroniti, \& Katz (1988),
which again assumed the equilibrium between gas and radiation,
and was improved by Okuda (2002).
Eggum et al. (1988) showed that mass accretion onto the black
hole occurs at a super-critical rate, although
the mass-accretion as well as mass-outflow rate were 
still variable in their simulations; that is, quasi-steady state
had not been achieved by their simulations.
This is the same also in the simulations by Okuda (2002), in which
the resultant luminosity slowly decreases with time.
He also found that the mass-accretion rate is sub-critical 
around the black hole, although the mass is injected from the 
outer boundary at a super-critical rate.
Note that the calculation times of these simulations were only
0.6 s and 1.6 s (in physical unit), respectively,
for the black hole mass of $10M_\odot$.  Notably, these
are shorter than the viscous timescale,
\begin{eqnarray}
  t_{\rm vis}\sim 5.7 {\rm s} \left(\frac{M}{10M_\odot}\right) 
  \left(\frac{r}{100r_{\rm g}}\right)^{3/2}\nonumber\\
  \times \left(\frac{\alpha}{0.1}\right)^{-1}
  \left(\frac{H/r}{0.5}\right)^{-2},
  \label{tvis}
\end{eqnarray}
with $\alpha$ being the viscosity parameter. 
More recently,
Okuda et al. (2005) performed long-term 2D-RHD calculations
of the super-critical accretion.
The luminosity as well as the mass-accretion rate
seems to be quasi-steady in their simulations, although
the flow structure is not steady yet.
Since the sum of the mass-accretion rate at the inner boundary 
and mass-outflow rate at the outer boundary is
much smaller than the mass-input rate at the disk boundary,
the mass within the computational domain continues to increase.
In addition, 
the photon-trapping did not appear in their simulations,
whereas the mass-accretion rate exceeds the critical value.

To summarize, despite the interesting simulations being made so far
by two groups
the quasi-steady structure of the super-critical accretion flows
still remains to be an open issue.
Further, interesting issues related to super-critical flow,
including the photon-trapping effects, 
the dependence of the observed luminosity on various viewing angles,
and the effects of metalicity on the flow structure,
have not been investigated previously.  This is a motivation
of the present study.

Here, we report for the first time
the quasi-steady structure of the super-critical disk accretion flows 
around black holes, which were revealed by 2D-RHD simulations.
Through the present simulations we mainly aim at understanding 
the dynamics of the viscous flow in the vicinity of the black 
hole of $r\lsim 100r_{\rm g}$.
Basic equations and our model are explained in \S 2, and
the numerical methods are described in \S3.
We will then display the quasi-steady flow structure and study
the photon trapping effects in the simulated flow in \S 4.  
Finally, \S 5 and \S6 are devoted to discussion and conclusions.

\section{BASIC EQUATIONS AND ASSUMPTIONS}
We solve the full set of RHD equations including the viscosity term.
We use spherical polar coordinates $(r, \theta, \varphi)$,
where $r$ is the radial distance, 
$\theta$ is the polar angle,
and $\varphi$ is the azimuthal angle.
In the present study, 
we assume the flow to be non self-gravitating,
reflection symmetric relative to the equatorial plane 
(with $\theta = \pi/2$), and axisymmetric with respect 
to the rotation axis (i.e., $\partial/\partial \varphi=0$).
We describe the gravitational field of the black hole in terms of 
pseudo-Newtonian hydrodynamics, in which the gravitational potential
is given by $\Psi(r)=-GM/(r-r_s)$ as was 
introduced by Paczynski \& Wiita (1980).
The basic equations are the continuity equation,
\begin{equation}
  \frac{\partial \rho}{\partial t}
  + \nabla \cdot (\rho {\bm v}) = 0,
  \label{mass_con}
\end{equation}
the equations of motion,
\begin{eqnarray}
  \frac{\partial (\rho v_r)}{\partial t}
  &+& \nabla \cdot (\rho v_r {\bm v}) 
  = - \frac{\partial p}{\partial r} \nonumber\\
  &+& \rho \left\{ 
    \frac{v_\theta^2}{r} + \frac{v_\varphi^2}{r}
    -\frac{GM}{(r-r_s)^2}
  \right\}
  + f_r
  + q_r,
  \label{mom_r}
\end{eqnarray}
\begin{equation}
  \frac{\partial (\rho r v_\theta)}{\partial t}
  + \nabla \cdot (\rho r v_\theta {\bm v}) 
  = - \frac{\partial p}{\partial \theta}
  + \rho v_\varphi^2 \cot\theta
  + r f_\theta
  + r q_\theta,
  \label{mom_th}
\end{equation}
\begin{equation}
  \frac{\partial (\rho r \sin\theta\cdot v_\varphi)}{\partial t}
  + \nabla \cdot (\rho r \sin\theta\cdot v_\varphi {\bm v}) 
  = r \sin\theta \cdot q_\varphi,
  \label{mom_varphi}
\end{equation}
the energy equation of the gas,
\begin{equation}
  \frac{\partial e }{\partial t}
  + \nabla\cdot(e {\bm v}) 
  = -p\nabla\cdot{\bm v} -4\pi \kappa B 
  + c\kappa  E_0 + \Phi_{\rm vis},
  \label{gase}
\end{equation}
and the energy equation of the radiation,
\begin{equation}
  \frac{\partial E_0}{\partial t}
  + \nabla\cdot(E_0 {\bm v}) 
  = -\nabla\cdot{\bm F_0} -\nabla{\bm v}:{\bm {\rm P}_0}
  + 4\pi \kappa B - c\kappa E_0.
  \label{rade}
\end{equation}
Here, $\rho$ is the gas mass density,
$\bm{v}=(v_r, v_\theta, v_\varphi)$ is the velocity,
$p$ is the gas pressure,
$e$ is the internal energy density of the gas,
$B$ is the blackbody intensity,
$E_0$ is the radiation energy density,
${\bm F}_0$ is the radiation flux,
${\bm {\rm P}}_0$ is the radiation pressure tensor,
$\kappa$ is the absorption opacity,
${\bm q}=(q_r, q_\theta, q_\varphi)$ is the viscous force,
and $\Phi_{\rm vis}$ is the viscous dissipative function,
respectively.
The radiation force, ${\bm f}_{\rm rad} = (f_r, f_\theta)$, 
is given by 
\begin{equation}
  {\bm f}_{\rm rad} = \frac {\chi}{c} {\bm F}_0,
  \label{frad}
\end{equation}
where $\chi (= \kappa+\rho \sigma_{\rm T}/m_{\rm p})$ 
is the total opacity 
with $\sigma_{\rm T}$ being the Thomson scattering cross-section
and $m_{\rm p}$ being the proton mass.

As the equation of state, we use
\begin{equation}
  p=(\gamma-1)e,
\end{equation}
where $\gamma$ is the specific heat ratio.
The temperature of the gas, $T$, can then be calculated from
\begin{equation}
  p=\frac{\rho k T}{\mu m_{\rm p}},
\end{equation}
where $k_{\rm B}$ is the Boltzmann constant and
$\mu$ is the mean molecular weight, respectively.

To complete the set of equations we apply
flux limited diffusion (FLD) approximation
developed by Levermore and Pomraning (1981).
In this framework,
the radiation flux is written as
\begin{equation}
  {\bm F}_0 = -\frac{c\lambda}{\chi}\nabla E_0,
\end{equation}
with the flux limiter, $\lambda$,
and the radiation pressure tensor is expressed
in terms of the radiation energy density via
\begin{equation}
  {\bm {\rm P}}_0 = {\bm {\rm f}} E_0,
\end{equation}
where ${\bm {\rm f}}$ is the Eddington tensor.
Here, the flux limiter is given by 
\begin{equation}
  \lambda = \frac{2+{\cal R}}{6+3{\cal R}+{\cal R}^2},
\end{equation}
with using the dimensionless quantity, 
${\cal R}=\left| \nabla E_0 \right| / \left( \chi E_0 \right)$.
The components of the Eddington tensor are
\begin{equation}
  {\bm {\rm f}} = \frac{1}{2}(1-f){\bm {\rm I}}
  +\frac{1}{2} (3f-1) {\bm n}{\bm n},
\end{equation}
where $f$ is the Eddington factor,
\begin{equation}
  f = \lambda + \lambda^2 {\cal R}^2,
\end{equation}
and ${\bm n}$ is the unit vector in the direction of the radiation
energy density gradient,
\begin{equation}
  {\bm n} = \frac{\nabla E_0}{\left| \nabla E_0 \right|}.
\end{equation}
This approximation holds
both in the optically thick and thin regimes.
In the optically thick limit, we find
$\lambda \rightarrow 1/3$ and $f \rightarrow 1/3$
because of ${\cal R} \rightarrow 0$.
In the optically thin limit of ${\cal R}\rightarrow \infty$,
on the other hand, 
we have $| F_0 | = cE_0$.
These give correct relations in the optically thick diffusion limit
and optically thin streaming limit, respectively.

For the absorption opacity,
we consider the free-free absorption, $\kappa_{\rm ff}$, 
and bound-free absorption, $\kappa_{\rm bf}$,
\begin{equation}
  \kappa=\kappa_{\rm ff}+\kappa_{\rm bf},
\end{equation}
where $\kappa_{\rm ff}$ and $\kappa_{\rm bf}$
are given by
\begin{equation}
  \kappa_{\rm ff} = 1.7\times 10^{-25} T^{-7/2}
  \left(\frac{\rho}{m_{\rm p}}\right)^2 \rm cm^{-1},
\end{equation}
(Rybicki \& Lightman 1979), and 
\begin{equation}
  \kappa_{\rm bf} = 4.8\times 10^{-24} T^{-7/2}
  \left(\frac{\rho}{m_{\rm p}}\right)^2 
  \left( \frac{Z}{Z_\sun} \right) \rm cm^{-1},
\end{equation}
with $Z$ being the metalicity
(Hayashi, Hoshi, \& Sugimito 1962),  respectively.

Here, we assume that only the $r\varphi$-component 
of the viscous stress tensor,
which plays important roles for
the transport of the angular momentum
and heating of the disk plasma,
is non zero,
\begin{equation}
  \tau_{r\varphi}=\eta r \frac{\partial}{\partial r}
  \left(\frac{v_\varphi}{r}\right),
\end{equation}
in the present study,
where $\eta$ is the dynamical viscosity coefficient.
Then, the radial and polar components of the viscous force are null
($q_r=q_\theta=0$),
and the right hand side of equation (\ref{mom_varphi}) 
is described as
\begin{equation}
  r \sin\theta\cdot q_\varphi = \frac{1}{r^2}\frac{\partial}{\partial r}
  \left( r^3 \sin\theta\cdot \tau_{r\varphi} \right).
\end{equation}
The viscous dissipative function is given by
\begin{equation}
  \Phi_{\rm vis}=\eta 
  \left[ r \frac{\partial}{\partial r}
  \left(\frac{v_\varphi}{r}\right)
  \right]^2.
\end{equation}
Finally, we prescribe the dynamical viscosity coefficient
as a function of the pressure
\begin{equation}
  \eta = \alpha \frac{p+\lambda E_0}{\Omega_{\rm K}},
\end{equation}
where $\Omega_{\rm K}$ is the Keplerian angular speed.
It is modified $\alpha$ prescription of the viscosity,
which is proposed by Shakura \& Sunyaev (1973). 
In this form, 
the dynamical viscosity coefficient is proportional to the
total pressure because of $\lambda \rightarrow 1/3$ 
in the optically thick regime.
In the optically thin regime, by contrast,
we find $\eta =\alpha p/\Omega_{\rm K}$ (or kinematic viscosity is
$\nu \equiv\eta/\rho = \alpha c_{\rm s}^2/\Omega_{\rm K}$
with $c_{\rm s} \equiv \sqrt{p/\rho}$ being isothermal sound velocity),
since $\lambda$ vanishes in this limit.

Here, we need to remark that 
the realistic formalism about the viscosity should be investigated
from the magneto-hydrodynamical point of view,
since the dominant sources of viscosity
would be chaotic magnetic fields and turbulence in the gas flow
(e.g., Machida, Hayashi, \& Matsumoto 2000; Stone \& Pringle 2001).

\section{Numerical Methods}
\subsection{The code}
We numerically solve the set of RHD equations 
shown in the previous section
by using an explicit-implicit finite difference scheme on the Eulerian grids.
Our methods and boundary conditions are similar to those of Okuda (2002),
but we adopt a different initial condition
and have carried out long-term calculations in order to examine
a quasi-steady structure of the supercritical accretion flows.
Since we assume the axisymmetry as well as the reflection symmetry,
the computational domain can be restricted to one quadrant
of the meridional plane.
The domain consists of spherical shells of 
$r_{\rm in} \leq r \leq r_{\rm out}$ and $0 \leq \theta \leq \pi/2$,
where $r_{\rm in}$ and $r_{\rm out}$ are the radial coordinates of the 
inner and outer boundaries of the computational domain, respectively,
and is divided into $N_r \times N_\theta$ grid cells,
amounting to $96\times 96$ meshes.
The $N_r$ grid points in radial direction are equally spaced logarithmically,
while the $N_\theta$ grids are equally distributed in such a way to achieve
$\Delta \cos\theta=1/N_\theta$.
All the physical quantities are defined at each cell center.

We divide the numerical procedure for the finite-difference equations
into the following steps:
[1] hydrodynamical terms for ideal fluid, 
[2] advection term in the energy equation of the radiation,
[3] radiation flux term in the radiation energy equation,
[4] gas-radiation interaction terms 
in the energy equations of the radiation and gas,
and 
[5] viscous terms in the momentum equation and 
energy equations of the gas.
Steps [1] and [2] are solved with the explicit method
while the steps [3], [4], and [5] are treated based on the implicit method.
In the first step, we use the computational hydrodynamical code,
the Virginia Hydrodynamics One.
It is based on the Piecewise Parabolic Method 
(Colella \& Woodward 1984). 
The equations (\ref{mass_con})-(\ref{gase}) 
except the viscous terms and
the gas-radiation interaction terms of equation (\ref{gase})
are solved in this step.
In the second step,
an integral formulation is used to generate a conservative
differencing scheme for the advection term of the equation (\ref{rade}).
The energy transport by the radiation flux term
is solved in the third step.
The radiation energy density is updated again 
with using the BiCGSTAB method for a matrix inversion.
In the fourth step, we solve the gas-radiation interaction terms
in equation (\ref{gase}).
All the terms on the right hand side of equation (\ref{rade}) 
except for the radiative flux term is also treated in this step. 
The radiation energy and gas energy is advanced simultaneously.
The method used in this step is basically the same as 
that described by Turner \& Stone (2001).
In the final step, we updates the azimuthal component of the velocity
by solving the viscosity term 
in the equation (\ref{mom_varphi}) with 
the Gauss-Jordan elimination for a matrix inversion.
The viscous dissipative function is also calculated in this step.

Throughout the present study, 
we assume $M=10M_\sun$, $\alpha=0.1$,
$\gamma=5/3$, and $\mu=0.5$.
The size of the computational domain is set to be
$(r_{\rm in}, r_{\rm out})=(3r_{\rm g}, 500r_{\rm g})$.
[Here, it is noted that our conclusion does not change so much,
even if we employ $r_{\rm in}=2r_{\rm g}$.]

\subsection{Boundary and Initial Conditions}
We employ the absorbing inner boundary condition so that 
the density, gas pressure, and velocity are smoothly dumped
(i.e. Kato, Mineshige, \& Shibata 2004).
Here, it is noted that our simulation results are not sensitive 
to the inner boundary condition nor to the location of the inner boundary.
The results does not change even if we use the free boundary conditions,
where all the matter and waves can transmit freely. 
The radiation flux is set to be $F_0^r=-cE_0$ at the inner boundary;
 that is, we apply
the condition of the optically thin limit at the inner boundary.

The outer boundary at $r=r_{\rm out}$ is divided into two parts:
the disk part ($\theta \geq 0.45\pi$) and the part above the disk
($\theta < 0.45\pi$).
Through the outer-disk boundary we assume that matter is 
continuously injected into the computational domain.
We set the injected mass-accretion rate (mass-input rate)
so as to be constant at the boundary.
In the present study, 
we explore the cases of $\dot{m}_{\rm input}=300$, $1000$, and $3000$,
where $\dot{m}_{\rm input}$ is 
the mass-input rate 
normalized by the critical mass-accretion rate.
The injected matter is supposed to rotate with sub-Keplerian speed,
and it has a specific angular momentum corresponding to the 
Keplerian angular momentum at $r=100r_{\rm g}$, since our main 
purpose of the present study is to investigate the viscous accretion 
flows within $r\lsim 100r_{\rm g}$. By setting the Keplerian 
radius ($100r_{\rm g}$) to be much smaller than the radius of the 
outer boundary, we can prevent that the outer boundary conditions
directly affect the accretion flow within $100 r_{\rm g}$.  With 
such a large outer boundary we can reproduce complex motions like 
circulation around $100r_{\rm g}$ in the disk; that is, the matter can 
transiently go out across the radius of $100 r_{\rm g}$ and return. 
At the outer boundary region above the accretion disk
we use free boundary conditions and allow for matter to go out 
but not to come in.
If the radial velocity is negative at the outermost grid,
it is automatically set to be zero.
We also employ the radiation flux in the optically thin limit,
$F_0^r=cE_0$, at the outer boundary.

With respect to the the rotation axis we assume $\rho$, $p$, $v_r$, and $E_0$
to be symmetric, while $v_\theta$ and $v_\varphi$ are antisymmetric. 
On the equatorial plane, on the other hand, 
$\rho$, $p$, $v_r$, $v_\varphi$, and $E_0$ 
are symmetric and $v_\theta$ is antisymmetric.

We start the calculations with a hot, rarefied, and 
optically-thin atmosphere. 
There is no cold dense disk in the computational domain, initially.
The initial atmosphere is constructed to approximately achieve
hydrostatic equilibrium in the radial direction; namely, 
its density profile is given by 
\begin{equation}
  \rho = \rho_{\rm out} 
  \exp \left[ \frac{\mu m_{\rm p}GM}{k_{\rm B}Tr_{\rm out}} 
  \left(\frac{r_{\rm out}}{r}-1 \right)
  \right],
\end{equation}
where $\rho_{\rm out}$ is the density at the outer boundary.
We employ $\rho_{\rm out}=10^{-17}{\rm g \, cm^{-3}}$
and $T=10^{11} \rm K$ in the present study.
Since this atmospheric gas is finally ejected out of the computational domain,
it does not affect the resultant quasi-steady structure.

\subsection{Time Step}
The time step is restricted by the 
Courant-Friedrichs-Levi condition.
We set the time step as
\begin{equation}
  \Delta t = \xi \min \left[ 
    \frac{\Delta r}{|v_r|+c_{\rm s}}, 
    \frac{r\Delta\theta}{|v_\theta|+c_{\rm s}}
  \right],
\end{equation}
where $\xi$ is a parameter, and
$\Delta r$ and $\Delta\theta$ are 
the cell sizes in the radial and polar directions, respectively.
Although we only show the results for the cases with $\xi=0.4$,
we also simulated the cases with $\xi=0.1$ or $0.05$,
confirming that our conclusions do not alter by this change.

\section{RESULTS}
\subsection{Quasi-steady Structure}
We first represent time evolution of 2D-RHD simulations.
Overall evolution is divided into two distinct phases:
the accumulation phase and the quasi-steady phase.

\begin{figure}[t]
\epsscale{1.18}
\plotone{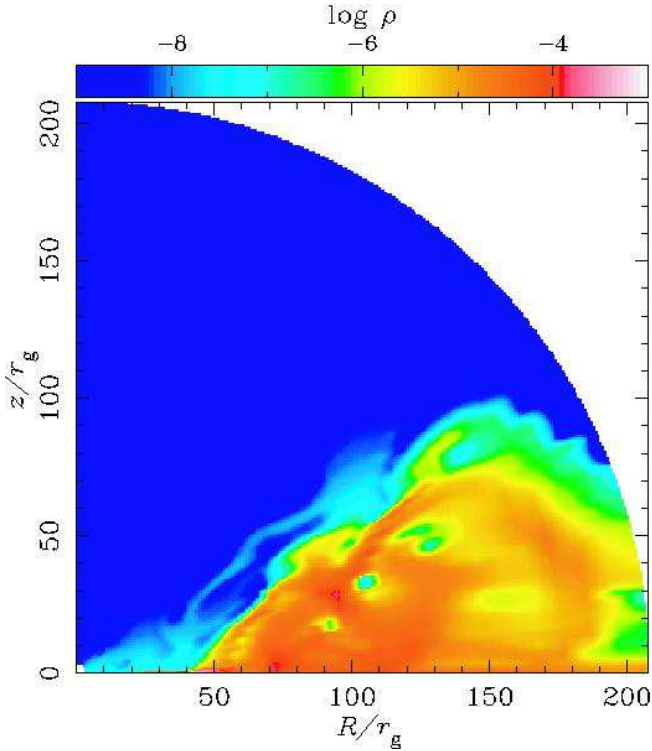}
\caption{
The 2D density distribution at the elapsed time of 
$t = $5.9 s for $M = 10 M_\odot$.
The adopted parameters are ${\dot m}_{\rm input}=1000$ and $Z=Z_\odot$.
\label{Phase1}
}
\end{figure}

The mass is continuously injected through the outer-disk boundary,
and creates continuous gas inflow because of the gravity force
by the central black hole.
Since angular momentum of the injected mass is set to be 
equal to the Keplerian angular momentum at $r=100r_{\rm g}$,
it is natural that the gas tends to accumulate
around the regions with the radius of $100r_{\rm g}$ by degrees
(see Figure \ref{Phase1}).
This is the accumulation phase.  Since
Eggum et al. (1988) and Okuda (2002) started calculations
with a cold dense disk, this phase did not appear 
in their simulations.
Eventually, the viscosity starts to work so that
the angular momentum of the gas can be transported outward,
which drives inflow gas motion in a quasi-steady fashion.
This is the quasi-steady phase.
%
%

\begin{figure}[h]
\epsscale{1.18}
\plotone{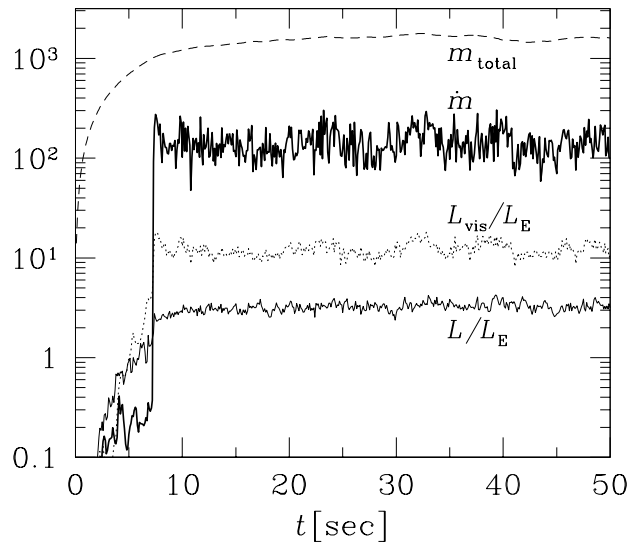}
\caption{
Time evolutions of the total mass within the computational domain
(dashed curve),
the mass-accretion rate onto the black hole normalized by $L_{\rm E}/c^2$
(thick solid curve),
the luminosity normalized by $L_{\rm E}$ (thin solid curve), 
and the total viscous heating rate normalized by $L_{\rm E}$ (dotted curve),
respectively, for the case with 
${\dot m}_{\rm input} = 1000$ and $Z = Z_\odot$.
\label{evolve}
}
\end{figure}

Such a two-step evolution is clear in Figure \ref{evolve}.
In this figure, we show the time evolution of normalized mass-accretion rate
onto the central black hole, 
$\dot{m}\equiv \dot{M}/(L_{\rm E}/c^2)$ (thick solid curve),
the luminosity, $L/L_{\rm E}$ (thin solid curve),
the viscous heating rate, $L_{\rm vis}/L_{\rm E}$ (dotted curve), and
the total mass contained within the computational domain, 
$m_{\rm total}\equiv M_{\rm total}/10^{19}\rm g\,cm^{-3}$ (dashed curve).
Here, we set the mass-input rate of $\dot{m}_{\rm input}=1000$ 
and the metalicity of $Z=Z_\odot$.
Both of the accretion rate and luminosity steadily increases until
$t\lsim 7$ s.  We also notice that
the mass-accretion rate is much smaller than the 
mass-input rate ($\dot{m}_{\rm input}=1000$), and that
the total mass within the computational domain rapidly increases 
with time, indicating mass accumulation really occurring in this phase.
Then starts the quasi-steady accretion phase (at $t\gsim 7$s), when
the mass-accretion rate exceeds critical value, $\dot{m}>1$,
and all the physical quantities stay nearly constant. 
[The constant $m_{\rm total}$ implies that 
the sum of the mass-accretion rate at the inner boundary 
and mass-outflow rate at the outer boundary is
equal to the mass-input rate.]
We can thus conclude that we could for the first time
succeed in reproducing the quasi-steady state
of the super-critical accretion flows with 2D-RHD simulations.
The critical time ($\sim$ 7s) separating these two phases
roughly coincides with the viscous timescale [equation (\ref{tvis})].
It is natural that the quasi-steady structure is obtained within
10 sec, which is the viscous timescale at the Keplerian radius
($r\sim 100r_{\rm g}$), 
although the size of the computational domain ($500r_{\rm g}$) 
is much larger. This is because that the injected matter accretes 
from the outer boundary to the Keplerian radius with the free-fall 
velocity so that the evolutionary timescale of the outer regions 
is given by the free-fall timescale at $r=500r_{\rm g}$,
$1.6 s \left(r/500r_{\rm g}\right)^{3/2} (M/10M_\odot)$.  This is
certainly shorter than the viscous timescale at $r=100r_{\rm g}$.
%
%

%
%
\begin{figure*}[t]
\epsscale{0.95}
\plotone{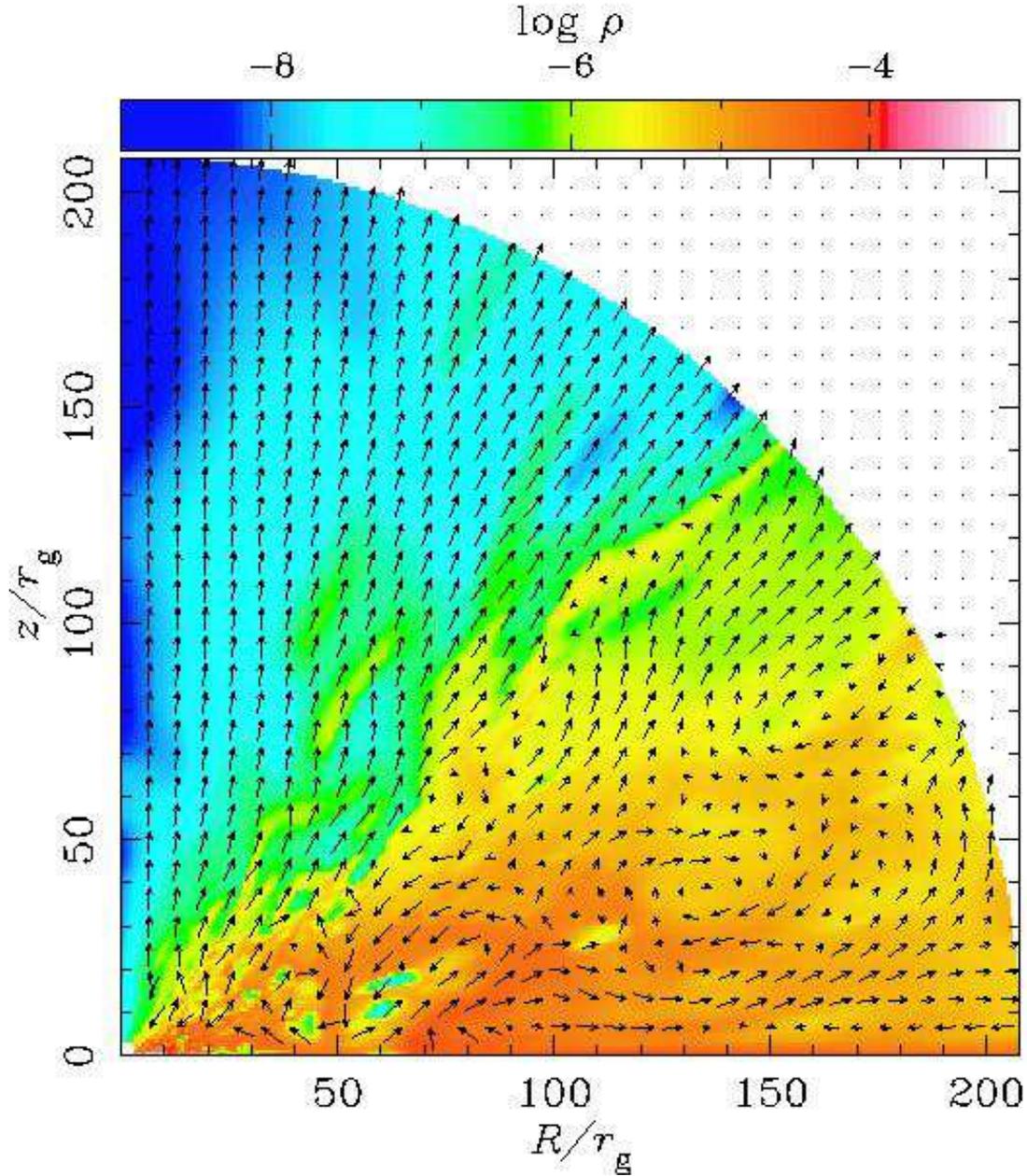}
\caption{
The 2D density distribution overlaid with the velocity vectors
at the elapsed time of $t = $10 s for $M = 10 M_\odot$.
The adopted parameters are ${\dot m}_{\rm input}=1000$ and $Z=Z_\odot$.
\label{Rho}
}
\end{figure*}
Let us, next, examine the quasi-steady structure in some details.
Figure \ref{Rho} displays the cross-sectional view of
the density distributions (with colors), overlaid with the
velocity vectors (with arrows) at $t=10\,\rm s$.  Here,
$x$- and $y$-axis are $R=r \sin \theta$ and $z=r \cos \theta$,
respectively. 
We understand with this figure that the flow structure is roughly
divided into two regions: the disk region around
the equatorial plane (characterized by orange color)
 and the outflow region above the inflow region
(characterized by blue color).
Roughly, the boundary is at $z/R \sim 0.8$; that is
the disk is geometrically and optically thick, as was
predicted by the slim-disk model.
However, the density distribution definitely deviates from 
that of the slim-disk model, since it is neither smooth nor 
plane parallel in the vertical direction.
We can even see a number of cavities in this figure.
The flow pattern is also complex, 
though the slim-disk model predicts the simple convergence flow.
We found the prominent circular motion within the disk
and the strong outflow which is generated at the disk surface.
The patchy structure around the boundary between the dense inflow region
and the rarefied outflow region seems to be caused by the 
Kelvin-Helmholtz (K-H) instability (discussed later).
The less dense gas in the outflow region penetrates into the disk body 
because of the K-H instability, thus forming the cavities.

\begin{figure*}[t]
\epsscale{0.95}
\plotone{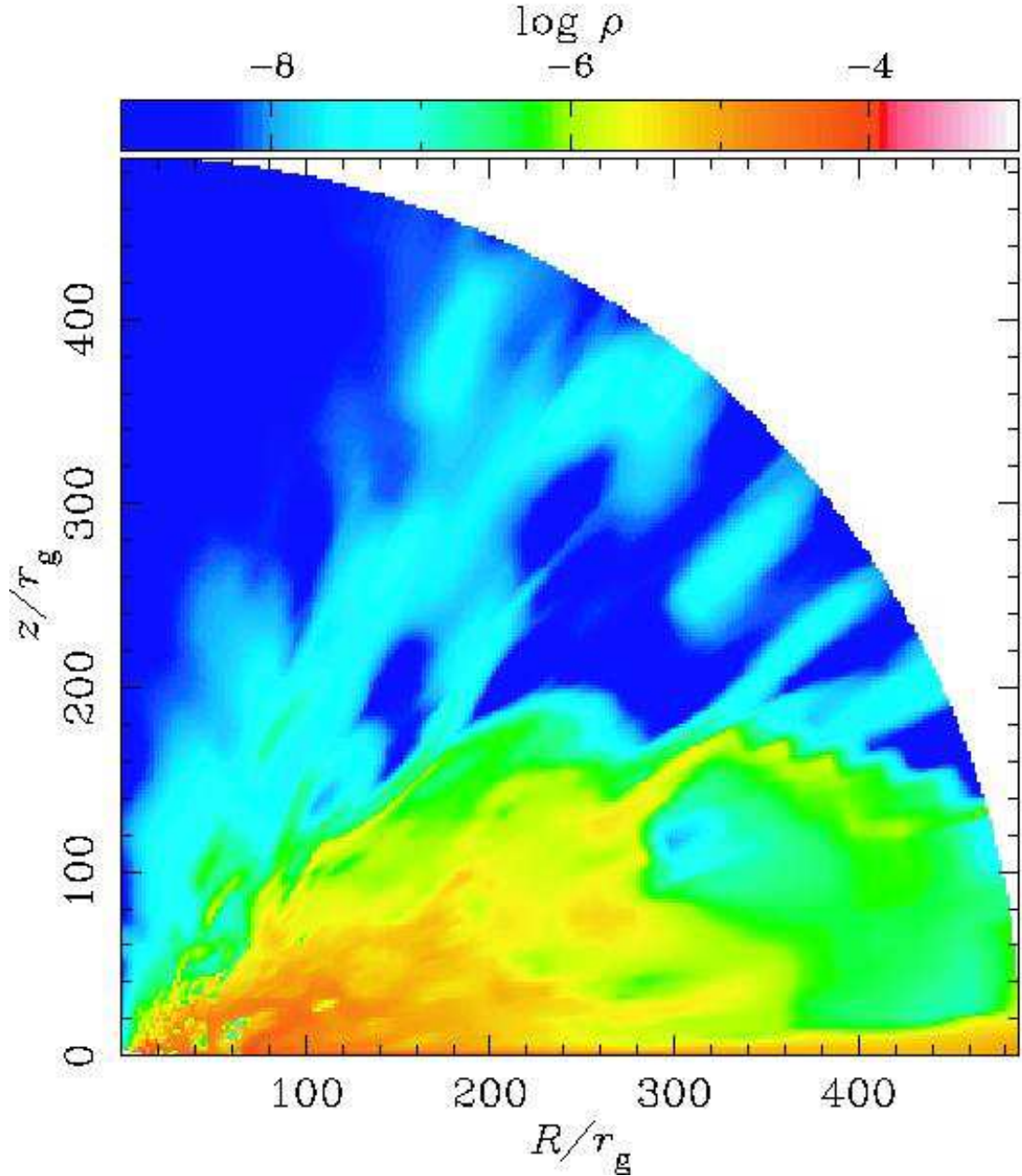}
\caption{
The 2D density distribution in the whole computational domain
at the elapsed time of $t = $10 s for $M = 10 M_\odot$.
The adopted parameters are ${\dot m}_{\rm input}=1000$ and $Z=Z_\odot$.
\label{ALL_Rho}
}
\end{figure*}
To understand the dynamics of our entire calculation domain,
we show the 2D density distribution in the whole region in Figure 
\ref{ALL_Rho}. 
We see from this figure that the density is larger around 
the equatorial plane ($|z| \lsim 10r_{\rm g}$) than at the 
large altitudes ($|z| \gsim 10r_{\rm g}$) in the outer region, 
$R\gsim 100r_{\rm g}$. This reflects that the injected mass accretes
along the equatorial plane. On the other hand, the viscous accretion 
disk forms inside, $r\lsim 100r_{\rm g}$. 
We focus the viscous accretion flows in the present study. The 
situation is apparently similar to that studied by Chakrabarti (1996), 
although they were concerned with subcritical flow and their study 
was not multi-dimensional nor radiation-hydrodynamical simulations.

\begin{figure*}[t]
\epsscale{1.1}
\plotone{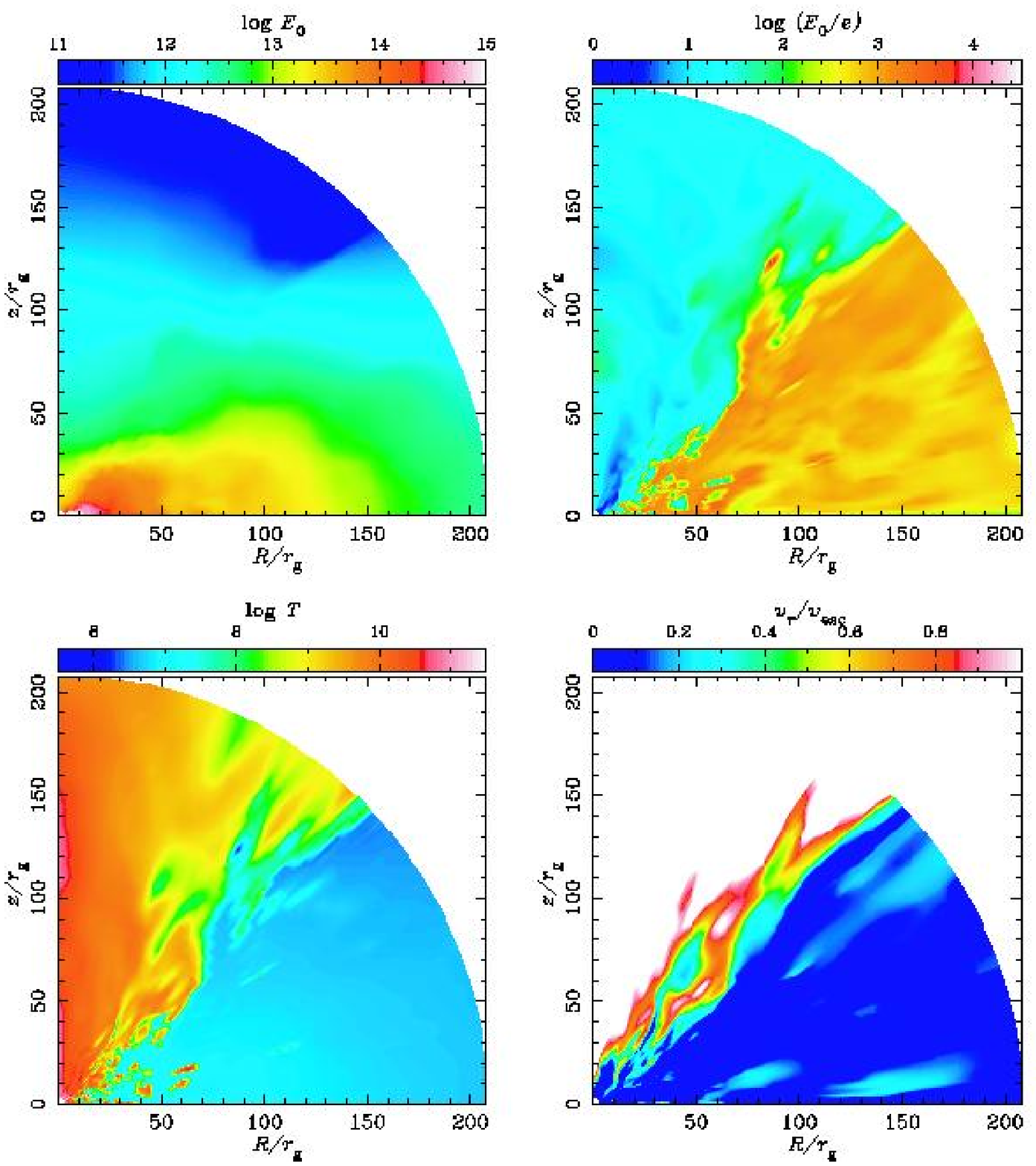}
\caption{
The 2D distribution of 
the radiation energy density (upper left),
the ratio of the radiation energy to the inertial energy of the gas
(upper right),
the gas temperature (lower left),
and the radial velocities normalized by the escape velocity (lower right),
respectively, at $t = 10$ s.
The adopted parameters are ${\dot m}_{\rm input}=1000$ and $Z=Z_\odot$.
\label{Center}
}
\end{figure*}
Figure \ref{Center} show the 2D distributions of 
radiation energy density (upper left), 
the ratio of the radiation energy to the internal energy of gas
(upper right),
gas temperature (lower left), 
and the radial velocity normalized by the escape velocity (lower right),
respectively, on the $R-z$ plane.
As shown in the upper left panel, the radiation energy distribution 
roughly coincides with the gas density distribution.  That is,
the radiation energy tends to be larger around the equatorial plane 
than that around the rotation axis.
Since the radiation energy distribution is smoothed 
due to the radiative diffusion within the disk,
there is no cavity found in the radiation energy distribution,
which makes a marked difference from
the density distribution (see Figure \ref{Rho}).

Radiation energy greatly exceeds the gas energy 
in the entire region, including the outflow region
in our simulations
(see the upper right panel).
This confirms that 
most of the regions is the radiation-pressure dominated and 
strong radiation pressure supports
the geometrically thick disk and drives the outflow.

The gas temperature distribution shown in the lower left panel
shows relatively low temperatures in the disk region, compared
with those in the outflow region, although
the viscous heating rate is much larger in the former than
in the latter.  This can be understood, because
radiative cooling is more effective in dense region
as a consequence of its strong density dependence.
The gas in the disk region is heated up by the viscous heating, and
the processed energy is effectively converted to the radiation energy.
Therefore, gas temperature does not rise so much within the disk.
Conversely, the gas cannot emit effectively in the outflow region,
since both free-free emissivity and bound-free emissivity
are more sensitive to the density than the gas temperature, 
$\propto \rho^2 T^{1/2}$.
It can be understood by the comparison between the lower left
and the upper right panels.
The radiation temperature ($\propto E_0^{1/4}$) in the outflow region
is lower than that in the disk region,
on the contrary to the gas temperature profile.

The lower right panel indicates the radial velocity normalized 
by the escape velocity.
The gas moves toward the black hole or 
flows outward slowly in the disk regions (see the blue area).
The white color indicates that the velocity exceeds the escape velocity
in this area.
It is found that the gas is accelerated through the radiation pressure
and is blown away to a large distance (see also the upper right panel).
Such a flow component will be identified as a strong disk wind.
Note that the outflow presented here is distinct in nature from 
that known as a bounce jet (Chen et al. 1997),
which arises because of a bounce of free-falling low angular momentum
material, when it goes through the centrifugal barrier at small $r$.

The outflow will also produce large absorption in the emergent spectra.
We also notice strong velocity shear at the boundary between
the disk region and the outflow region.
This complex density profile around the disk surface as 
shown in Figure \ref{Rho} 
is explained as a consequence of the K-H instability.

The growth timescale of the K-H instability is roughly given by
\begin{equation}
  t_{\rm KH} \approx 
  \frac{1}{k v_r}
  \left( \frac{\rho_{\rm disk}}{\rho_{\rm out}} \right)^{1/2},
\end{equation}
where $k$ is the wavenumber,
while $\rho_{\rm disk}/\rho_{\rm out}$ 
is the density ratio of the disk region to the outflow region
at the disk boundary.
Here, we assume the incompressible fluid as well as
$\rho_{\rm disk}\gg\rho_{\rm out}$ and neglect the gravity.
Also, the viscosity is not considered,
since $r\theta$-component of the viscous stress tensor
is set to be zero in the present simulations.
By setting $k=2\pi/10r_{\rm g}$, $v_r = 0.1c$,
and $\rho_{\rm disk}/\rho_{\rm out}=10$,
we find $t_{\rm KH} \sim 5\times 10^{-3}$ s.
Note that this is shorter than the escape time,
$r/v_r=0.1$ s, for $r=100r_{\rm g}$ and $v_r=0.1c$.
That is, there is ample time for the K-H instability
to grow before the material flows outward.

\begin{figure}[b]
\epsscale{1.18}
\plotone{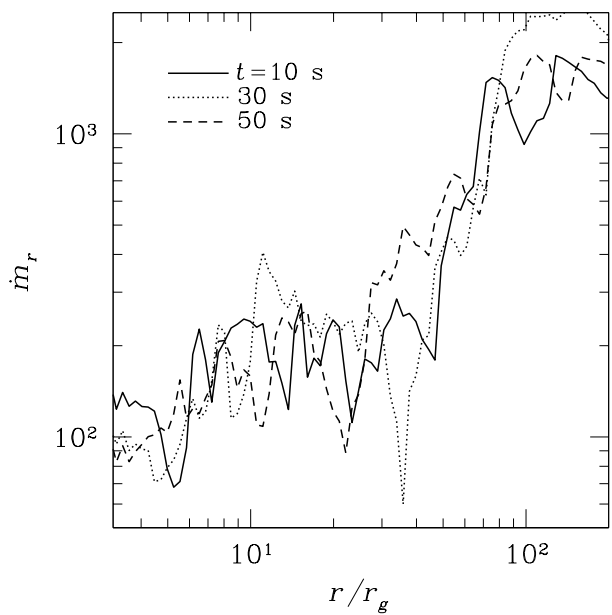}
\caption{
The mass-accretion rates as functions of the radius
at $t = $10 s (by the solid line), 30 s (dotted line), 
and 50 s (dashed curve).
The adopted parameters are ${\dot m}_{\rm input}=1000$ and $Z=Z_\odot$.
\label{mdot}
}
\end{figure}

The mass-accretion rates as a function of the radius 
are displayed in Figure \ref{mdot} for the case of
$\dot{m}_{\rm input}=1000$ and $Z=Z_\odot$.
The solid, dotted, and dashed curves 
indicate the $\dot m$ profiles at elapsed times of
 $t=10$ s, 30 s, and 50 s, respectively.
Here, the mass-accretion rate at each radius is evaluated by 
\begin{equation}
  \dot{m}_r=- \frac{c^2}{L_{\rm E}} 
  \int 2\pi r^2 \rho(r,\theta) \min[0,v_r(r,\theta)] 
  \sin\theta d\theta.
\end{equation}
It is found that the mass-accretion rates 
are not constant in the radial direction but decrease inward,
as the flow approaches the black hole.
Roughly, we find $\dot{m}_r \propto r$.  In addition, we find that
the radial $\dot{m}_r$ profile remains nearly the same after 7s, from
which we can conclude that the flow is in a quasi-steady state.
This $\dot{m}_r$ change is caused by a cooperation between
a wind mass loss around the disk surface and the circular motion 
deep inside the disk.
Note that accretion rates are assumed to be constant in space
in the slim-disk model formulation.

The multi-dimensional numerical simulations of RIAFs have 
shown that the accretion disks are convectively unstable, thereby the 
circular motion being driven within the flow (e.g., Igumenshchev 
\& Abramowicz 1999; Stone, Pringle, \& Begelman 1999; McKinney 
\& Gammie 2002). The convection might cooperate with the K-H 
instability in generation of the circular motion as well as 
the cavities found in our simulations. The simulations of RIAFs have 
also revealed that $\dot{m}_r$ increases with radius as 
$\dot{m}_r \propto r^{3/4-1}$, which is similar to our value,
and was attributed to the circular motion as well as the bipolar 
outflows (Stone, Pringle, \& Begelman 1999,
see also Narayan, Igumenshchev, \& Abramowicz 2000 for 
self-similar solution). Here, we need to 
emphasize that although the flow structures look similar at first 
glance, the basic physical processes are distinct from those of RIAF 
simulations.  That is, entropy is carried by fluid in the RIAF, 
whereas it is mostly by photons in the present case. In addition, 
since the RIAF simulations do not basically take into account of 
the radiative cooling, they would overestimate the driving force 
of the outflows.

\subsection{Photon-trapping}
The photon-trapping characterizes the super-critical accretion flows.
It works to reduce the energy conversion efficiency,
$L/\dot{M}c^2$ (see, e.g. Paper I; Ohsuga et al. 2003).
As a result, the flow luminosity becomes insensitive to 
the mass-accretion rate when $\dot m \gg 1$.
However, simple stream lines were assumed in the previous studies
on super-critical flows, including the slim-disk model.
Even if we do not consider possible multi-dimensional gas motions,
photon trapping works to some degree, leading to a reduction in
the energy conversion efficiency.
Such multi-dimensional gas motion and associated reaction of the radiation
can be calculated in the present RHD simulations, 
since we have calculated the radiation energy transport,
being coupled with the gas dynamics.

\begin{figure}[h]
\epsscale{1.18}
\plotone{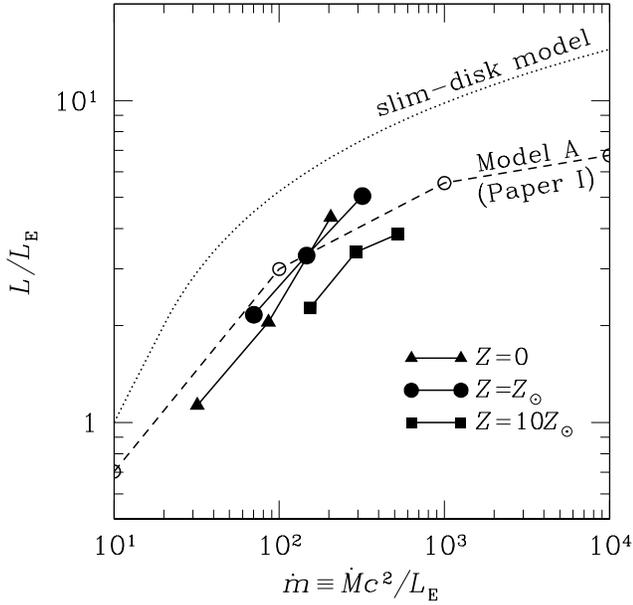}
\caption{
The luminosities as functions of the mass-accretion rate
onto the black hole.
The symbols of triangles, circles, and squares indicate the metalicity
of 0, $Z_\odot$, and 10 $Z_\odot$, respectively.
The normalized accretion rate of each symbol is ${\dot m}_{\rm input} =$ 
10, 100, and 1000 from the lower-left to the upper right, respectively.
For comparison, the same but based on the slim-disk model
and that based on Model A of Paper I are plotted
by the dotted and dashed curves, respectively.
The present results agree more with the latter than the former.
\label{trap}
}
\end{figure}

Figure \ref{trap} plots the luminosity as a function of the
mass-accretion rate onto the black hole.
The filled squares, circles, and triangles indicate the results
for $Z=10Z_\odot$, $Z_\odot$, and 0,
respectively, with different mass-input rates,
$\dot{m}_{\rm input}=300$, $1000$, and $3000$ from the left to the right.
We also indicated in the same figure the luminosity
calculated based on the slim-disk model (Watarai et al. 2000)
and the one based on Model A of Paper I
with dashed and dotted curves, respectively.
In Paper I, a simple model for the accretion flow is employed,
and the luminosity 
with carefully taking account of the photon-trapping
is calculated 
by solving energy transport inside the accretion flows.
[More precisely, the luminosity plotted in Figure \ref{trap} is 
 the corrected one, which fixes initial small errors in
 their Model A (see Figure 1 in Paper I).]
Here, it should be stressed 
that the mass-accretion rate onto the black hole
is not input parameter but is calculated dynamically
in the present simulations, although it was a parameter in both 
of Model A of Paper I and the slim-disk model.
The resultant $\dot m$ profile was shown in Figure \ref{mdot}.

It is evident in Figure \ref{trap} that
the calculated luminosity agrees more with Model A of
Paper I, rather than the slim-disk model, 
in all the cases.
This proves that the two-dimensional effects of photon-trapping 
is really significant in the super-critical accretion flows.
This result is, in a sense, surprising. 
In Model A in Paper 1, 
we assumed that the viscous heating occurs only 
in the vicinity of the equatorial plane,
although the gas might be heated up at the high altitude.
Since the photons emitted at deep inside the disk
tends to be more effectively trapped in the flow,
we anticipated that 
the photon-trapping effects would be reduced
in the realistic situation, compared with Model A.

We also argued in Paper I
that photon trapping effects may be attenuated by the presence of
large-scale circulation motion, which could help photon diffusion
motion and thus considerably reduce photon traveling time to 
the surface of the flow.
However, our current results show significant photon trapping effects
even when we explicitly include complex flow motions. 

Figure \ref{trap} also shows that the mass-accretion rate 
onto the black hole increases with an increase of the metalicity
(for fixed mass-input rates).
In our simulations the gas with higher metalicity 
has larger absorption opacities so that the gas energy 
can be more effectively converted into the radiation energy,
yielding smaller gas pressure than in the metal-poor case.
The gas could be effectively blown away by the strong gas pressure.
However, there is a counter effect.
The gas with large absorption opacity enhances the radiation energy.
Enhanced radiation flux and large opacity can drive
strong radiation-pressure driven outflows.
The physical reason will be investigated in future.

\begin{figure}[h]
\epsscale{1.18}
\plotone{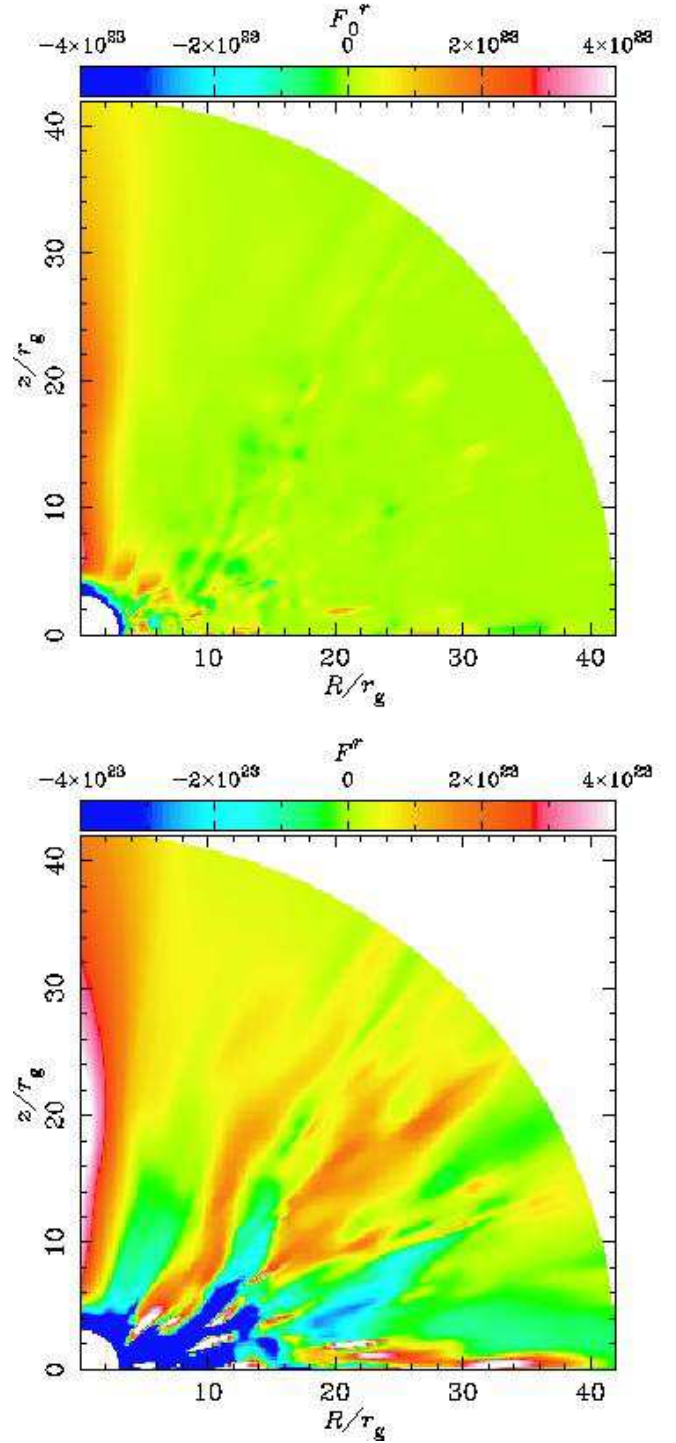}
\caption{
The 2D distribution of the radial component of the
radiation energy flux in the comoving frame (upper) 
and that in the inertial frame (lower), respectively.  
The adopted parameters are ${\dot m}_{\rm input}=1000$ and $Z=Z_\odot$.
The difference between them indicate the contribution of
advective energy transport, $v_r E_0$ (see text).
\label{Flux}
}
\end{figure}
Let us see more explicitly how significant photon trapping is.
Figure \ref{Flux} compares the distributions of
the radial component of the radiative flux in the comoving frame (upper panel)
and that in the inertial frame (lower panel) at $t=10\,\rm s$.
Other parameters are the same;
the mass-accretion rate is $\dot{m}_{\rm input}=1000$ and 
the metalicity is $Z=Z_\odot$.
The former radiative flux ($F^r_0$) 
is roughly proportional to the radial gradient of 
the radiation energy distribution.
On the other hand, the latter flux ($F^r$) includes 
the advective transport of radiation energy ($v_r E_0$)
in addition to the former flux; namely, 
we approximately have $F^r \sim F^r_0 + v_r E_0$.
In other words, differences between two panels
represent how significant photon trapping is.
In fact, we see a significant difference between the two panels;
whereas the blue area (indicating large radiation flux) is restricted
to the vicinity of the black hole in the upper panel,
it is more widely spread over the entire disk region in the lower panel.
This is a direct manifestation of the photon-trapping effects.

%
%
\begin{figure*}[t]
\epsscale{1.1}
\plotone{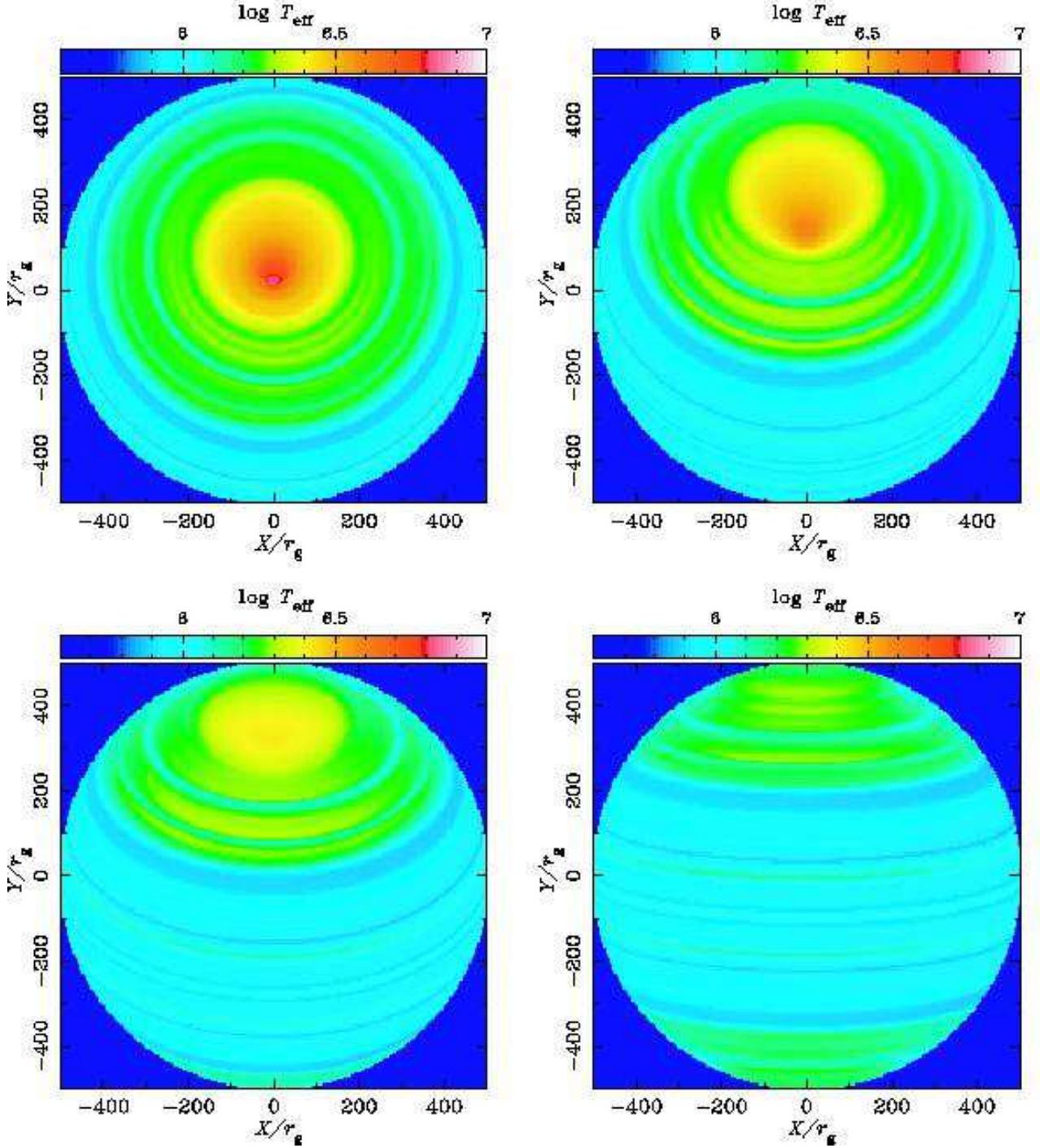}
\caption{
Surface temperature images for the model for 
different viewing angles: $\cos i=1/8$, $3/8$, $5/8$, and $7/8$
in the upper-left, upper-right, lower-left, and the lower-right panels,
respectively.
The origin of the Cartesian coordinate 
on the observer's screen, $(X,Y)=(0,0)$, 
is set at the location of the black hole.
The black hole at the center
The adopted parameters are ${\dot m}_{\rm input}=1000$ and $Z=Z_\odot$.
\label{image}
}
\end{figure*}
\subsection{Collimated Emission}
Since the super-critical accretion flows are 
geometrically and optically thick,
the observed images and luminosity should strongly 
depends on the viewing (inclination) angle.
We calculate the intensity map
with using the monochromatic radiation transfer equation,
\begin{eqnarray}
  \bm{l}\cdot\nabla I
  &=& \rho 
  \Gamma^3   
  \left(1+\frac{\bm{v}\cdot \bm{l}_0}{c} \right)^3 \nonumber\\
  &\times &
  \left[ \kappa B + \frac{c}{4\pi}\frac{\rho\sigma_{\rm T}}{m_{\rm p}}E_0
    -\chi I_0 \right],
\end{eqnarray}
where $I$ is the specific intensity, 
$\Gamma\equiv(1-v^2/c^2)^{-1/2}$ is the Lorentz factor,
$\bm{l}=(\sin\Theta \cos\Phi,$ $\sin\Theta \sin\Phi, \cos\Theta)$ and
$\bm{l}_0=\Gamma^{-1} (1-\bm{v}\cdot\bm{l}/c)^{-1}
\{\bm{l}-(\Gamma/c)\bm{v}+(\Gamma-1/\bm{v}^2)(\bm{l}\cdot\bm{v})\bm{v} \}$
are the directional cosine
in the inertial and comoving frame, respectively,
with $\Theta$ and $\Phi$ being the azimuthal and 
polar angles of radiation propagation 
in the inertial frame.
Here, we assumed isotropic scattering.

The calculated effective temperature ($T_{\rm eff}$) maps at $t=50\,\rm s$ 
are shown in Figure \ref{image} for various viewing angles:
$\cos i=1/8$, $3/8$, $5/8$, and $7/8$.
The other parameters are
$\dot{m}_{\rm input}=1000$ and $Z=Z_\odot$.
On the observer's screen, the black hole lies at the center
of the Cartesian coordinate $(X,Y)$.
Since we use the spherical computational domain
with the size of $500 r_{\rm g}$ in the present study,
the blue four corners of the maps indicate outside of the domain.
In the face-on view (see the upper left panel; $\cos i = 7/8$),
the effective temperature exceeds $3\times 10^6$K 
within $100 r_{\rm g}$
and amounts $\sim 10^7$K in the central region.
Such a high temperature region disappears in the upper right panel
(the case with $\cos i = 3/8$).
In this panel, 
the most luminous region is found on the upper side (not at the center)
and is elongated in the vertical direction.
These are caused by an occultation of the innermost part of
the flow by the outer parts (see Fukue 2000, Watarai et al. 2005
for self-occultation effects based on the slim-disk model).
Since the accretion flow is both geometrically and optically thick,
the innermost region cannot be seen for large inclination angles, $i$.

\begin{figure}[b]
\epsscale{1.18}
\plotone{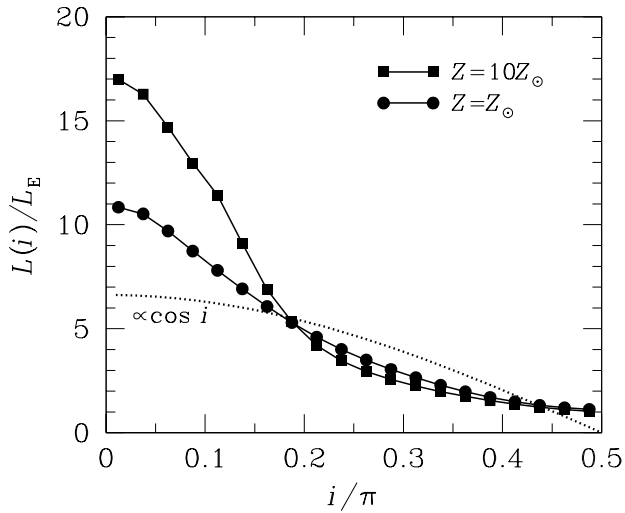}
\caption{
Isotropic luminosities as functions of the viewing angle, $i$,
for the cases with $Z = Z_\odot$ and 10 $Z_\odot$.
The dashed line represents the curve of $\cos i$;
i.e., the variation expected if the flow shape is flat and
geometrically thin (infinitesimally).
\label{Lth}
}
\end{figure}

The emission from the super-critical accretion flows is 
mildly collimated due to the same reason.
Figure \ref{Lth} represents the viewing angle-dependence of the
isotropic luminosity (normalized by bolometric luminosity)
for $Z=Z_\odot$ (circles) and $Z=10Z_\odot$ (squares), respectively.
The other parameters are
$t=50$ s and $\dot{m}_{\rm input}=1000$.
Here, the isotropic luminosity is calculated 
by assuming isotropic radiation field,
$L(i)=4\pi D^2 I\left( i \right)\zeta$,
with $D$ being the distance from an observer
and $\zeta$ being a revised factor.
The luminosity calculated by solving radiation transfer equation
does not always coincide with that evaluated by the FLD method.
Here, we assume $\zeta=1.7$ to fit these luminosities.
As expected, 
the isotropic luminosity is quite sensitive to the viewing angle,
indicating that the emission from the flow is mildly collimated.
If the flow shape would be flat and (geometrically) thin,
and if no collimation would occur,
the observed luminosity should vary along the cosine curve,
as is indicated by the dotted curve.

We also find that the angle dependence of the isotropic luminosity 
is more enhanced in large-metalicity case of $Z=10Z_\odot$
than that of $Z=Z_\odot$.
This is because of different density distribution.
As have already mentioned,
the mass-outflow rate is smaller in the high metalicity case
than in the low metalicity case.
As a result, the density contrast between the disk region
and the outflow region above the disk gets larger
with an increase of the metalicity.
Accordingly, the hot innermost region of the high metal flows
are easier to be observed owing to smaller optical depth 
in the outflow region for small viewing angles.

The super-critical accretion flows would be identified as 
very high $L/L_{\rm E}$ objects in the face-on case,
since the emission is mildly collimated in the polar direction
and the bolometric luminosity itself exceeds the Eddington luminosity.
The former effect is enhanced in the case of high metalicity
(large absorption opacity).
It is also true that super-critical flow may be identified as
fairly sub-critical source, if the viewing angle is large
(Watarai et al. 2005).

\section{DISCUSSION}
\subsection{Comparison with the slim-disk model}
\subsubsection{Flow structure}
\begin{figure}[t]
\epsscale{1.18}
\plotone{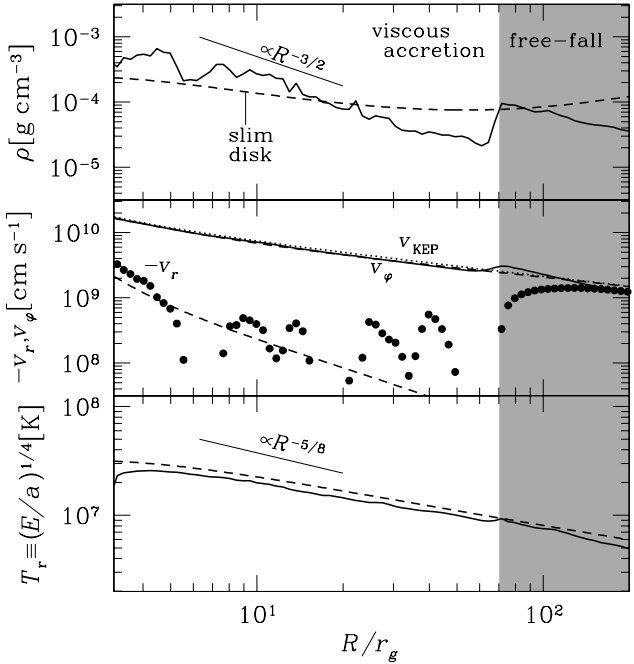}
\caption{
The density profile (top), 
radial and rotation velocity profiles (middle),
and radiation temperature profile (bottom)
at the equatorial plane, $z=0$,
averaged in the elapsed time between $t=40$ s and 50 s.
The adopted parameters are ${\dot m}_{\rm input}=1000$ and $Z=Z_\odot$.
Shaded region is the free-falling region, where the injected matter
falls at the free-fall velocity, whereas the viscous accretion disk
forms around the black hole, $r\leq 70r_{\rm g}$.
The dashed curves indicate the profiles of one-dimensional numerical 
study of the slim-disk model (Watarai et al. 2000).
The thin solid lines in the top and bottom panel
means the slope of the self-similar solution of the slim-disk model.
The dotted curve in the middle panel is 
the Keplerian velocity.
\label{slim}
}
\end{figure}

Here, we directly compare the structure of 
the super-critical accretion flows based on our 2D simulations
with that of the slim-disk model.
Figure \ref{slim} represents the time-average density profile (top panel),
radial and rotation velocity profiles (middle panel),
and radiative temperature profile (bottom panel) 
on the equatorial plane in the quasi-steady accretion phase of $t=40-50$ s.
The adopted parameters are ${\dot m}_{\rm input}=1000$ and $Z=Z_\odot$.
The resultant profiles of the one-dimensional numerical study 
for the slim-disk model are represented by the dashed curves
(Watarai et al. 2000).
As have already mentioned in section 4.1, the injected matter
falls with the free-fall velocity outside the Keplerian radius 
(shaded region). 
A viscous accretion disk forms in the inner part (white region).
The free-falling matter penetrates slightly within the Keplerian 
radius, thus a separation radius appears around 70 $r_{\rm g}$.
We focus on the viscous accretion regime in this subsection.

The density profile is shown by the solid curve in the top panel.
The density peak around $70r_{\rm g}$ is made by accumulation of 
the injected matter, where occurs because of the centrifugal barrier.
We compare the density profile in the viscous accretion regime
with that of numerical and self-similar solutions of the slim-disk model.
In this panel, it is found that the slope of the density profile 
of our simulations is steeper than that of the numerical solution 
of the slim-disk model.
We also find that the slope 
is close to that of the self-similar solution of the 
slim disk, $\rho \propto R^{-3/2}$ (Watarai et al. 1999,
see also Spruit et al. 1987),
whereas the profile becomes flatter in the vicinity of the 
black hole ($R<5r_{\rm g})$.
Here, we note that the slope in our simulations is steeper than 
that of the RIAFs. The RIAF simulations showed the density profile 
to be $\rho \propto r^{-1/2}$ (McKinney \& Gammie 2002).

In the middle panel, 
the radial and rotation velocities are plotted by the filled circles
and the solid curve.
The dotted curve indicates the Keplerian velocity,
$v_{\rm KEP}=\sqrt{GMR}/(R-r_{\rm g})$.
It is found that the rotation velocity 
nicely agrees with the Keplerian velocity
as well as the result of the numerical solution of the slim disk.
However, the radial velocity profile is not smooth and largely 
deviates from that of the slim-disk value,
whose slope agrees with that of the Keplerian velocity (dotted curve) 
in the free-fall regime.
The radial velocity remarkably varies at $R=5-70~r_{\rm g}$,
and the radial velocity sometimes becomes negative,
around $R=6$, $20$, and $60~r_{\rm g}$.
Such a complex $v_r$ profile is formed by the prominent circular motions
around the equatorial plane (see Figure \ref{Rho}).

The bottom panel shows the radiation temperature profile,
where the radiation temperature is defined as 
$T_{\rm r} \equiv (E/a)^{1/4}$ with
$a$ being the radiation constant.
As shown in this panel, 
the profile 
nicely agrees with that of the one-dimensional numerical solution and
is flatter than that of 
the self-similar solution of the slim-disk model, 
$T_{\rm r}\propto R^{-5/8}$.

To sum up, we can thus conclude that 
the quasi-steady structure of the super-critical accretion flows
are apparently similar to but, precisely speaking,
deviates from the prediction of the slim-disk model.

\subsubsection{Effective temperature}
\begin{figure}[t]
\epsscale{1.18}
\plotone{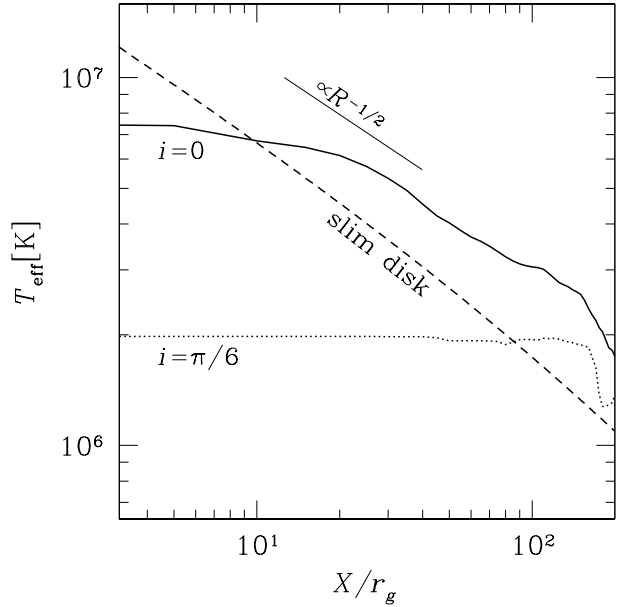}
\caption{
The effective temperature profiles 
in the $X$-directions at $Y=0$
for the case with $i=0$ (solid curve) and $\pi/6$ (dotted curve).
Here, $(X,Y)$ is the Cartesian coordinate on the observer's screen, 
and the black hole lies at the center.
The other parameters are ${\dot m}_{\rm input} = 1000$, $Z = Z_\odot$,
and $t=50$ s.
The dashed curve indicates the result of the one-dimensional numerical 
study of the slim-disk model (Watarai et al. 2000),
and the thin solid line means the slope of 
the self-similar solution of the slim-disk model.
\label{slim_Teff}
}
\end{figure}

Next, we represent again the effective temperature profile
in order to directly compare with the prediction of the slim-disk model,
although 
the 2D images of the effective temperature have already shown 
in Figure \ref{image}.
Figure \ref{slim_Teff} represents the effective temperatures 
at $Y=0$ as a function of $X$ for $i=0$ and $\pi/6$,
where $X$ and $Y$ are the horizontal and vertical coordinates 
on the observer's screen (see Figure \ref{image}).
The other parameters are ${\dot m}_{\rm input} = 1000$, $Z = Z_\odot$,
and $t=50$ s.
It is found that 
the slope of the effective temperature profile for $i=0$
becomes flatter than that of the slim-disk model, 
$T_{\rm eff}\propto R^{-1/2}$, within $30r_{\rm g}$.
In contrast, we found that the profile is consistent 
with the slim-disk model in the outer region.
As have already mentioned,
the effective temperature as well as the luminosity is
quite sensitive to the viewing angle,
since the accretion flow is both geometrically and optically thick.
We found that the effective temperature profile for $i=\pi/6$ 
is almost flat.

The slim-disk model succeeds in reproducing the
observed behavior of Ultra-luminous X-ray sources (ULXs) 
and narrow-line Seyfert 1 galaxies (NLS1s),
which are thought to be candidates of near- or super-critical flows
(Watarai, Mizuno, \& Mineshige 2001; Mineshige et al. 2000;
Kawaguchi 2003).
The SEDs are produced based on the blackbody emission 
with the effective temperature coupled with the 
some modification.  We, here, note that the temperature
profile obtained by our current simulations show some
deviations from that of the slim-disk model.

\subsection{Blueshifted absorption by outflow matter}
One characteristic feature of the super-critical accretion flow
is its generation of the radiation-pressure driven wind.
Our simulations show that a large quantity of gas is blown away by 
the strong radiation pressure above and below the accretion disk.
Such outflow material, which is thought to be highly ionized 
by the radiation from the accretion flow,
would absorb the continuum emission.
Therefore, the super-critical accretion objects have 
the blueshifted absorption lines by the highly ionized ions.
Such blueshifted absorption in high-ionization lines
was observed in the UV spectra of broad absorption line quasars
(Weymann, Carswell, \& Smith 1981; Becker et al. 2000). 
Moreover, Pounds et al. (2003a, 2003b) and Reeves, O'Brien, \& Ward (2003)
reported by the blueshifted X-ray absorption lines
that the highly ionized matter with outflow velocity
of order $\sim 0.1c$ in quasars,
and the flow is likely to be optically thick to the
electron scattering within $r\sim 100r_{\rm g}$.
In our simulations, the typical outflow velocity is $0.1c$ and 
the optical depth of the outflow regions is 
around $1.2 (\rho/10^{-8} {\rm g\,cm^{-3}}) (r/100r_{\rm g})$
(This estimated value does not depend on the black-hole mass 
because of $\rho\propto M$ and $r\propto M^{-1}$.)
These are remarkably close to the observed results, and therefore,
our simulations basically account for the origins of high-velocity
outflows.  Detailed comparison with the observations is
left as future work.

\subsection{Growth of supermassive black holes}
We reveal by 2D-RHD simulations
that the black hole can swallow the gas 
at the super-critical accretion rate, 
although the luminosity exceeds the Eddington luminosity.
This result implies that the black hole can rapidly grow,
and the growth timescale is given by
$M/\dot{M} = 4.5\times 10^6 (\dot{m}/100)^{-1} \rm yr$.
Thus, 
if the supermassive black holes in the galactic nuclei
are built up by the super-critical accretion,
such a growing phase would be quite short
because of a short growth timescale.

Umemura (2001) suggested the radiation-hydrodynamical model
for the formation of the supermassive black holes,
in which the supermassive black holes grow in the 
ultraluminous infrared galaxies (ULIRGs) due to 
the mass accretion caused by the radiation drag.
Based on this model,
Kawakatu, Umemura, \& Mori (2003) also claim that 
proto-quasars, which have relatively small black hole
and show only narrow emission-lines,
appear just after the ULIRG phase, $t\sim 10^{8-9}$ yr.
However, such proto-quasar phase might not be observed 
if the black hole grows at super-critical accretion rate
in ULIRG phase.
A seed black hole in the ULIRG 
can evolve to a supermassive black hole
via the super-critical accretion by the end of the ULIRG phase,
since the mass-accretion rate from the circumnuclear regions 
can exceed the critical rate
due to the effective radiation drag.
Thus,
we presumably observe the grown-up quasars after the ULIRG phase,
and proto-quasars are obscured by huge amount of the dust
in the ULIRG.

As shown in Figure \ref{mdot},
our simulations show that the black hole can 
effectively grow in the case that the accreting gas is metal rich.
This implies that super-critical accretion flows
tend to emerge in the metal rich objects
like starburst galaxies or star-forming regions.
This tendency is also consistent with the observed results
that the NLS1s, 
which is though to be near- or super-critical accretion objects, 
are metal rich objects
(Nagao et al. 2002; Shemmer \& Netzer 2002; Shemmer et al. 2004).
It is proposed that NLS1s accrete at super-critical rates
based on the study of the optical band
(Kawaguchi 2003; Collin \& Kawaguchi 2004).

The growth of the black hole may gradually slow down,
since the density as well as the absorption opacity 
of the super-critical accretion flows 
are small around the massive black hole,
if the normalized mass-accretion rate does not change so much.
Observational constraints are proposed by Yu \& Tremaine (2002),
by which most of luminous quasars must be
sub-critical accretion phase.
They showed that the luminous quasars have the 
energy conversion efficiency of $L/\dot{M}c^2 \gsim 0.1$,
based on the study of the local black hole density
and the luminosity function of the quasars.
Since this efficiency is much smaller than 0.1 
in the super-critical accretion flows,
most of luminous quasars must be
sub-critical accretion phase (see also Soltan 1982).

\subsection{Future work}
\subsubsection{Spectral energy distribution}
Throughout the present study,
we use the frequency-integrated energy equation of the radiation
[see equation (\ref{rade})].
By solving the monochromatic radiation transfer equation,
we can calculate the effective temperature profiles 
and show them in Figure \ref{image}.
However, the emergent SEDs of the super-critical accretion flows
are not a simple superposition of blackbody spectra
with various effective temperatures (Ohsuga et al. 2003).
The photons generated deep inside the disk are difficult to reach
the disk surface and thus moves downward with gas.
Furthermore, most of photons generated in the vicinity of the black hole
will be immediately swallowed by the black hole and cannot 
contribute to the emergent SED.
Frequency dependent radiation-hydrodynamical simulations are necessary
to resolve this issue.

According to the current simulations the hot outflow
with temperature of $10^{9-10}$ K appears above the disk.
The Compton $y$-parameter of this outflow region is comparable to or 
larger than unity, meaning that
the inverse Compton scattering cannot be ignored.
Seed soft photons emitted from the disk region should be 
Compton up-scattered, producing high-energy, non-thermal emission
which contributes to the emergent SED.
The corona above the super-critical accretion disk 
has been claimed to fit the observed SEDs of NLS1s and 
very-high state of BHCs (Wang \& Netzer 2003; Kubota \& Done 2004),
although the formation mechanism of the corona is poorly known.
The hot outflow shown in our simulations might resolve this issue.

Here, we should note that
the Comptonization process is not included in our simulations.
If the Compton cooling were effective,
the hot outflow might be cooled to some extent.
The bulk Comptonization may also contributes
the production of the high-energy photons
because of large radial velocities, $v_r > 0.1c$, near the black hole.
The detailed study of the SED with Comptonization
is, however, beyond the scope of this paper and should be
explored in future work.

\subsubsection{Viscosity}
We assumed that only the $r\varphi$-component of the viscous stress tensor
is non-zero while all the other components vanish, because
the $r\varphi$-component plays an essential role of
in the angular momentum transport within the disk.
If the $r\theta$- as well as $\theta\varphi$-component 
were non zero, the structure near the disk surface might change.
Figure \ref{Rho} clearly shows abrupt velocity changes
across the boundary between the disk and the outflow regions,
which is thought to promote the K-H instability.
The growth of this instability might be suppressed by 
the $r\theta$- and $\theta\varphi$-components of viscosity.
These components might also partially suppress the circular motion
in the disk. Stone et al (1999) discovered by the simulations of 
RIAFs that the flow structure differs from that given by 
Igumenshchev \& Abramowicz (1999). It implies that the inclusion 
of the components of the viscous stress tensor suppresses the 
convection, since the $r\theta$-component is excluded in Stone 
et al. (1999).  This will be examined in future study.

More importantly, we need coupled MHD and RHD simulations,
since the source of disk viscosity is likely to be of magnetic origins
(Hawley, Balbus, \& Stone 2001; Machida, Matsumoto, \& Mineshige 2001; 
Balbus 2003 for a review).  
Recently, local radiation MHD simulations of the 
accretion flow have been performed by 
Turner et al. (2003) and Turner (2004).
We definitely need global simulations to include global field effects.

\subsubsection{The FLD approximation}
Finally, we need to comment on the FLD approximation.
It is known that the FLD approximation does not always give 
good results for the regions with moderate optical thickness.
The FLD approximation is thought to be valid 
in the disk region, which has a large optical depth,
but the spherical shape of the radiation energy density distribution
above the disk might be artificial.  

In addition,
the radiation drag force cannot be treated in the FLD approximation,
whereas it might play an important role for the transport of 
the angular momentum within the outflow region.
%
%
Since the timescale of the angular momentum transport via
the radiation drag, which is given by 
\begin{eqnarray}
  \frac{M_{\rm g}}{L/c^2}
   \sim 1.6{\rm s} \left(\frac{L}{4L_{\rm E}}\right)^{-1}
   \left(\frac{\rho}{10^{-8} {\rm g\,cm^{-3}}}\right) \nonumber\\
   \times
   \left(\frac{r}{200r_{\rm g}}\right)^3 
   \left(\frac{M}{10M_\odot}\right), 
\end{eqnarray}
(with $M_{\rm g}$ being the mass within the outflow region),
is about ten times longer than the escape time,
$r/v_r=0.2{\rm s}(r/200r_{\rm g})(v_r/0.1c)^{-1}(M/10M_\odot)$,
roughly ten percent of the angular momentum could be extracted
in the outflow region.
[The exact expressions for the radiation drag are found
in the literature
(Umemura, Fukue, \& Mineshige 1997; Fukue, Umemura, Mineshige 1997;
Ohsuga et al. 1999).]

Radiation drag arises where there exists a large velocity difference 
between radiation sources and the irradiated matter.
In the FLD approximation, however, 
the radiation flux is determined solely by the 
local gradient of the radiation energy, and 
photon trajectories cannot be properly considered.
Thus, the FLD approximation can not treat the radiation drag force,
in principle.
It would be better to solve the radiation transfer equations 
without using this approximation.

Begelman (2002) suggested that 
strong density inhomogeneities on scales much smaller 
than the disk scale height
is formed due to the photon bubble instability
in the radiation pressure-dominated accretion disks,
and the disk can remain geometrically thin even as the 
the maximum luminosity exceeds Eddington 
luminosity by a factor of one hundred.
However, the FLD approximation is not suitable method
for investigating the radiation fields 
in such inhomogeneous structure.
The detailed study of the photon bubble instability
should be explored in future work.

\vspace{0.1cm}
\section{CONCLUSIONS}
By performing the 2D-RHD simulations,
we, for the first time, investigate the quasi-steady structure 
of the super-critical accretion flows around the black hole
with particular attention being paid on the photon-trapping effects.
We have obtained several new findings:

(1) 
The quasi-steady structure of the super-critical flow 
is divided into two parts: the disk region (with mostly inflow)
and the outflow region above the disk.
The two regions are separated by a sharp density jump.
The gas outflow driven by the strong radiation pressure is produced 
around the rotation axis.
Further, there exists velocity shears at the boundary,
which causes K-H instability around the disk surface,
producing the patchy structure as well as the circular motion
within the disk region.  
Convection may also be responsible
for such inhomogeneous structure.

(2) 
The photon-trapping plays an important role in the super-critical 
accretion regime.
The advective energy transport is substantial,
and the large amount of photons generated inside the disk
is swallowed by the black hole without radiated away.

(3)
Our 2D-RHD model shows some differences from the slim-disk model.
The slim-disk model assumes a simple convergence flow,
while our simulations revealed rather complex gas motion and structure.
We also found that the mass-accretion rate is not constant in space
but decreases as matter accretes, roughly as $\dot m \propto r$,
as a result of wind mass loss 
and circular motion.
The calculated luminosity of the flows agrees more with
the prediction of Paper I
rather than that of the slim-disk model.

(4)
The emission of the super-critical accretion flows is moderately collimated.
The apparent luminosity could become more than ten times larger than
the Eddington luminosity.
The super-critical accretion flows would be identified 
as high $L/L_{\rm E}$ objects in the face-on view,
but not, if the viewing angle is large, for which
self-occultation tends to reduce the total luminosity
and the maximum flow temperature.

(5)
The mass-accretion rate increases
with increase of the absorption opacity (metalicity) 
of the accreting matter.
It implies that the black hole 
tends to grow up faster in the metal rich regions as in
starburst galaxies or star-forming regions.
In addition, the growth of the black hole may gradually slow down,
since the density as well as the absorption opacity 
of the super-critical accretion flows 
are small around the massive black hole,
if the normalized mass-accretion rate is kept constant.

\acknowledgments

The authors would like to thank the anonymous referee 
for important comments and suggestions. 
The calculations were carried out at 
Department of Physics in Rikkyo University.
This work is supported in part by 
Research Fellowships of the Japan Society
for the Promotion of Science for Young Scientists, 02796 (KO),
by the Grant-in-Aid of the Ministry of Education,
Culture, Science, and Sports, 14740132, 
by the Promotion and Mutual Aid Corporation for Private Schools of Japan
(MM),
by the Grants-in-Aid of the
Ministry of Education, Science, Culture, and Sport, 
(14079205, 16340057),
and by a Grant-in-Aid for the 21st Century COE
 {\lq\lq}Center for Diversity and Universality in Physics" (SM).

\end{document}